\documentstyle[twocolumn,aps,prl]{revtex}
\begin{document}
\draft
\title{Phase Diagram of strongly correlated one-dimensional
 fermions in incommensurate potentials}
\author{ Juan Carlos Chaves\cite{email1} and Indubala I.
Satija\cite{email2} }
\address{
 Department of Physics and\\
 Institute for Computational Sciences and Informatics,\\
 George Mason University,\\
 Fairfax, VA 22030}
\author{ Mauro M. Doria}
\address{ Department of Physics\\
Institute de Fisica\\
Universidade Federal Fluminense\\
Campus da Praia Vermelha\\
Niteroi, 24210-340, RJ, Brazil}
\date{\today}
\maketitle
\begin{abstract}
We study the interplay between strong correlations  
and incommensurability on fermions
using mean field as well as
exact many-body Lanczos diagonalization techniques.
In a two-dimensional parameter space, mean field phase diagram of infinite
system shows
a $critical$ phase with multifractal characteristics 
sandwiched between a
Bloch-type extended and a localized phase.
Exact numerical computation on finite size systems
of the Kohn charge stiffness $D_c$, characterizing
transport properties, 
and the charge exponent $K_{\rho}$, characterizing
the superconducting pairing fluctuations,
shows the existence of a phase with 
superconducting correlations. 
This phase 
may be characterized
by a power law scaling of the charge stiffness constant
in contrast to the Anderson localized phase
where $D_c$ scales exponentially with the size of the system.
We argue that 
this intermediate phase may be a {\it dressed up} analog of the
$critical$ phase of the noninteracting fermions 
in an incommensurate potential.
\end{abstract}

\section{Introduction}
The study of metal-insulator transition in strongly correlated 
one-dimensional (1D)
disordered systems have been the subject of many 
recent studies.\cite{DisInt1,DisInt2}
These studies have been motivated by various different viewpoints.
A great deal of theoretical research in this area
has been to account for the magnitude of
persistent currents in mesoscopic rings.\cite{DisInt1,DisInt2}
Another motivation for these studies has been the argument
that 1D models in a staggered field
may describe the physics of two-dimensional systems; the subject of
great interest in high $T_c$ superconductivity.\cite{Campbell}

In this paper, we describe our study of the phase diagram of many-body spinless 
fermion chain in presence of a sinusoidal potential with periodicity which
introduces a new length scale in the problem.
We approach the problem of understanding
the phase diagram of this aperiodic system
from a perspective very different from the previous studies.
It has been known that noninteracting systems
in incommensurate potentials exhibit
metallic, insulating as well as an exotic fractal phase
which is in-between metallic and insulating and is usually referred
as $critical$.\cite{Sokoloff,Hiramoto92}
These systems which are in-between periodic and random
have attracted a great deal of attention due to the possibility
of exhibiting a novel type of transport
as the multifractal
wavefunctions are characterized by almost a power law decay. 
Technical complexities associated with studying strongly correlated systems
make the study of the effect of strong correlations on incommensurate
systems a rather difficult task.
Hence, the natural questions like 
how the metal-insulator transition and the fractal characteristics of non-interacting fermions
are affected when many-body fermion-fermion correlations
are taken into account has remained an open theoretical challenge.

We recently published our preliminary 
results\cite{JSK1}
on this problem using Lanczos diagonalization
method where we focussed
on the transport studies of 1D spinless model
in the presence of both repulsive as well as the attractive interaction.
One of the intriguing results of our study for 
attractive interaction
was the possibility of a new phase
where the charge stiffness may scale as a power law.
The purpose of this paper is three fold.
Firstly, we present our simulation results for systems of bigger
sizes
than those reported in our earlier paper. This strengthens the possibility
of a new type of phase in strongly correlated systems.
Secondly, we compute the charge exponent
$K_{\rho}$ which characterizes the long range correlations in the system.
Our results show that the regime with superconducting fluctuations overlaps
the regime where we see a new type of transport characteristics.
Thirdly, we describe self-consistent
mean field (MF) theory results where the mean field is assumed to be due to
copper pairs. The self-consistent mean field theory shows 
the existence of a critical phase, sandwiched between an extended and localized
phase. Interestingly, this critical phase  exists
 in a finite parameter space of measure
directly proportional to the strength of the cooper pair interactions. 
We correlate the MF and the exact diagonalization results and
argue that the new type of behavior seen in numerical simulation
may be a dressed up analog of the critical phase of 
noninteracting fermions. 

In section II,  we describe our model and two interesting limits where
the model has been extensively studied.
In section III, we discuss mean-field results for the
spinless fermion case. In addition to the various remarks in the introduction,
this section
will further bring to focus our
motivations for the proposed study.
In section IV, we describe our results for the charge exponent
$K_{\rho}$ and
charge stiffness $D_c$.
In section V we summarize our results.

\section{ Model System and its two interesting limits }

Consider an interacting spinless fermion model on a 1D ring in 
a quasiperiodic potential,
\begin{equation}
H = - \sum_{i=1}^{N} (c_i^\dagger c_{i+1} + c_{i+1}^\dagger c_i) +
V \sum_{i=1}^{N} n_i n_{i+1} + \sum_{i=1}^{N} h_i n_i.
\label{spinless}
\end{equation}
The $c_i$ and $n_i$ are respectively the fermion and number operators
at site $i$.
The site dependent potential is chosen to be of the form,
$h_i=\lambda \cos(2 \pi \sigma i)$. Here,
$\lambda$ represents the strength of the potential and $\sigma$
is an irrational number which is chosen for convenience to be
the {\it Golden Mean} ($\frac{\sqrt{5}-1}{2}$).
This parameter describes the two competing length scales in the system: namely,
the fermion lattice space and the periodicity of the sinusoidal potential.

The above model has two interesting limits that have been studied extensively.
For $\lambda=0$, the interacting spinless fermion problem could
be mapped to the Heisenberg-Ising XXZ spin
problem\cite{Jordan-Wigner} for which
a closed Bethe's ansatz solution exists.\cite{XXZ}
For all densities, the model exhibits
a Luttinger liquid phase.
The Luttinger liquid phase
differs from the conventional Fermi liquid which accounts for the
conduction behavior in conventional metallic phase. 
In Fermi liquid phase the low energy
excited states of the interacting electron gas can be
described in term of quasiparticle excitations which are analog
to the single particle excitations of a free Fermi system.
In contrast, the Luttinger phase 
is characterized by many-body collective excitations.
For attractive interaction ($V < 0 $), the model exhibit SC
correlations that can be described in terms of
the charge exponent $K_{\rho}$,
\begin{equation}
<O_iO_{i+r}> \propto r^{-{1/K_{\rho}}}
\end{equation}
where $O_i$ is a (singlet) pairing operator,
\begin{equation}
O_i=\frac{1}{\sqrt(2)} ( c_ic_{i+1} + h.c.)
\end{equation}
Therefore, $K_{\rho} > 1$ signals the existence of SC fluctuations.

Other interesting limit of the model is the $V=0$ case where the model describes
the famous Harper equation\cite{Harper55} which has also been studied
both numerically as well as more recently by Bethe's ansatz.\cite{BA}
The Harper equation is a paradigm in the study of low-dimensional
incommensurate systems as it
exhibits a metal-insulator transition
in one dimension.\cite{Sokoloff,Hiramoto92}
At the
onset of transition $\lambda_c =2$, the quantum states are neither
extended nor
localized but instead exhibit fractal character and have been termed
as {\it critical}.  The spectrum contains an
infinite number of gaps and is believed to be
a Cantor set of zero measure. These interesting aspects of the wave function
and the spectra
have been shown to be reflected in the transport properties such as
Launder resistance.\cite{Liu86,Cruz93}

The possibility of a phase which is neither Anderson insulating
nor metallic
in strongly correlated systems is a fascinating theoretical problem. We explore this
first using mean-field theory and then using exact many body
Lanczos 
diagonalization on finite size systems. In correlated systems,
we will focus 
directly on the transport properties characterized by the stiffness constant.
This is in contrast to the mean-field case where 
the single particle
wavefunctions can be used as the diagnostic tool 
for characterizing
the nature of the phase
in the system.

\section{ Mean Field Theory} 

For attractive interaction, we assume the existence of electron pairing
resulting in a mean field $\Delta$
\begin{equation}
\Delta = \frac{1}{N} \sum_{i=1}^{N} <c_i c_{i+1}> 
\end{equation}
In terms of the mean field $\Delta$, the Hamiltonian becomes
\begin{eqnarray}
H & \approx & -t \sum_{i=1}^{N}
c_i^\dagger c_{i+1} + c_{i+1}^\dagger c_i)
-V\Delta \sum_{i=1}^{N} (c_{i+1}^\dagger c_{i}^\dagger + c_i c_{i+1})\nonumber
\\ &-&
\sum_{i=1}^{N} h_i n_i
\end{eqnarray}
The parameter $\Delta$ is a function of both $t$ and $V$ and has to be
determined self-consistently.

The above mean-field Hamiltonian is the fermion representation
of spin-$\frac{1}{2}$ $XY$ chain in a transverse field.\cite{Lieb68}
The fermion operators $c_i$ are related to spin operators $S_i$
by Jordan Wigner
transformation,
\begin{equation}
S_n^+  = c_n^\dagger \exp \lgroup i\pi \sum_{m<n} c_m^\dagger c_m \rgroup
\end{equation}
\begin{equation}
S_n^-  =  \exp \lgroup -i\pi \sum_{m<n} c_m^\dagger c_m \rgroup c_n
\end{equation}
\begin{equation}
S_n^z  = c_n^\dagger c_n - \frac{1}{2}.
\end{equation}

The parameters $t$ and $V\Delta$ are related to the exchange interactions
in the spin model: $t$ is proportional to the sum of the exchange interaction
in the 2-dimensional spin space while the fermion non-conserving term $V\Delta$
is proportional to spin space anisotropy.
The $V$ dependent term in the Hamiltonian breaks the $O(2)$
symmetry of the spin chain and the spin model is in the universality
class of the Ising model with anisotropy proportional to $V\Delta$.
The mean-field $\Delta$ can be determined self-consistently by
the methods described by Lieb et al 
(for example as the results shown in fig. 1).\cite{Lieb68} 
In the absence of magnetic field ($\lambda = 0$), the $\Delta$ can be determined
analytically in the following two limits:
For $\frac{t}{V}=0$, 
$\Delta=\frac{1}{2\pi}$ and for
$t=\Delta V$, 
$\Delta=\frac{1}{8}$.

The anisotropic spin model in the presence of modulating magnetic field
was studied recently as the perturbed Harper
equation.\cite{KS} 
Numerical diagonalization 
as well as exact decimation scheme showed that
the model exhibits extended (E), localized (L) as well as a critical (C)
phase. A novel aspect of the model is the fact that unlike Harper, the C phase
exists in a finite parameter interval.
The term $V\Delta$ 
results in fattening the critical point of Harper to a 
critical phase, sandwiched between the Bloch phase and the localized
phase. The boundaries between the Bloch states and critical states
as well as the boundaries between the critical and localized states
were found numerically to be determined by simple relations:
\begin{itemize}
\item $\lambda < 2(t+V\Delta)$ , we have a {\bf E} phase.
\item If $\lambda > 2(t-V\Delta)$ , we have a {\bf L} phase.
\item If $2(t+V\Delta) \leq \lambda \leq 2(t+V\Delta)$ , 
we have a {\bf C} phase.
\end{itemize}
Therefore, by determining the parameter $\Delta$ 
self-consistently,
the phase boundaries of the model can be inferred.
It should be noted that the mean-field theory can also be done in the absence of cooper pairs
by writing mean field term as $<c_i^{\dagger}c_{i+1}>$. 
In this case, the MF Hamiltonian belongs
to the universality class of Harper equation which 
exhibit E-L transition. The critical phase
in this case reduces to a critical line. Such a MF theory gives a self-consistent solution
for both repulsive as well as attractive interaction $V$.

Figure 1 shows the mean field $\Delta$ obtained self-consistently.
It should be noted that the strength of the aperiodic field 
$\lambda$
determines a critical value of $V$ below which the mean field 
$\Delta$
is zero. For repulsive interaction, the self-consistent 
value of $\Delta$
was always found to be zero. 

Figure 2 shows the MF phase diagram in the two-dimensional parameter
space of $V-\lambda$. For attractive interaction, the model
exhibits E, L as well as fat C phase beyond a critical value of $\lambda$.
Exact renormalization study of the C phase\cite{KS} 
has shown that the wave function exhibits self-similarity.
The exponents characterizing the self-similar behavior at the onset of
E-C and C-L transition are different from those of the fat critical
phase sandwiched between these two phases.\cite{KS}
Another interesting
aspect of this diagram is the fact that C-L transition line coincides
with the magnetic transition to long range order which is in the
universality class of Ising model in transverse field. 

The MF model exhibits a gap in the energy spectrum
(except at the conformal point which corresponds to the transition to
magnetic long range order in the XY spin model).
However, we argue that the transition under consideration in
disordered systems is the Anderson transition and therefore
the transport properties are determined by the
nature of single particle wave function.

The Lanczos's numerical results shown in the next section confirm the
existence of superconducting pairing for attractive interaction.
We would like to argue that this justifies our basic assumption
about the existence of a mean field is derived from the cooper pairs.
Reentrant nature of the MF diagram may be just the artifact of
MF assumption. Therefore, the MF phase diagram may describe qualitatively
the physics of strongly correlated systems in the regime where the superconducting
fluctuations exist.
In spite of the limitations of MFT in 1D, the interesting phase diagram
provides a strong incentive for further exploration of phase diagram
with a view to check the possibility of an exotic phase
(with localization
properties intermediate between metallic 
and insulating) in 
strongly correlated systems. 

\section{ Lanczos Diagonalization Results}

Motivated by the possibility of a new type of phase, we next
use Lanczos exact diagonalization
method to obtain the stiffness constant $D_c$ which determines
the transport properties of the system.
In addition, we also compute the charge exponent to
check the validity of MF theory based on the assumption that the
MF derived from the fermion pairing.
As explained below, both these quantities can be determined from the ground
state properties of the system.

The first step in this computation is to determine the
ground state energy $E_0$
of a one-dimensional ring with $N$ lattice sites subjected
to a transverse magnetic flux $\Phi$. The
the Kohn stiffness constant $D_c$ is then given by,
\cite{DisInt1,Cstiffness}
\begin{equation}
D_c = \frac{N}{2} \frac{d^2 E_0(\Phi)}{d \Phi^2} |_{\Phi = \Phi_{min}}.
\end{equation}
The charge exponent
$K_{\rho}$ can also be determined
in terms of ground state properties by first computing the compressibility 
$\kappa$.
As explained by P. Prelovsek et all\cite{Campbell} the compressibility
for finite systems can be
calculated from values of ground state energies:\cite{Campbell}
\begin{equation}
\frac{1}{\rho^2 \kappa} = \frac{N}{4} [E_0(N_e+2) -2E_0(N_e) + E_0(N_e-2)]
\end{equation}
Here $N_e$ are the number of fermions in the system with fermion density
$\rho=N_e/N$.
Now the following two equations relate $\kappa$ to the the charge exponent
via the the renormalized
Fermi velocity $u_{\rho}$,
\begin{eqnarray}
K_{\rho} &=& \pi \frac{D_c}{u_{\rho}} \\
\frac{1}{\rho^2 \kappa} & = & \frac{\pi u_{\rho}}{2 K_{\rho}}
\end{eqnarray}

Our simulations are done for systems
of various sizes $N$ and
electronic densities $\rho = \frac{N_e}{N}$, in two-dimensional
parameter space
$V$ and $\lambda$.
To simulate golden mean quasiperiodicity into the model,
we used Lanczos methods
for systems of various Fibonacci sizes.
Furthermore, we worked with
densities  which are
the rational approximants to the
the square of the golden mean $\sigma^2$.
This procedure provides several possible sizes (5, 8, 13 and 21)
for which the Lanczos
diagonalization can be done keeping the density of the fermions
almost a constant. 
It should be noted that unlike the previous studies involving 
random disorder, we cannot work with arbitrary sizes. 
This limits not only the number of sizes that
we can study, but also forces us to work with densities 
different from half-filling .
Therefore, our studies are for systems away from half-filling
where the umklapp processes become irrelevant and the system
in absence
of disorder is metallic.
The next possible Fibonacci size $34$ is rather hard to simulate
using present day technology, and therefore,
our results are in fact for the maximum possible exact diagonalization
size (for simulating incommensurate effects) that can be done
with current regular computers.
To obtain additional data points, we also show some results for
densities which are rational approximants of 
$\frac{\sqrt 10-1}{3}$ such us $\rho = \frac{13}{18}$. However, the 
fermion density for this case
is different from those of Fibonacci size systems.

Figures 3 and 4 respectively
show our results for the charge exponent and the charge stiffness. 
Figures (a) and (b) correspond to two different densities and are
included here to show the consistency of our results independent
of the density.
As shown in figure 3, comparison with the
$\lambda=0$ case show that the presence of a new competing length
enhances the SC pairing fluctuations in the strongly correlated
fermions. This effect becomes
specially strong around $V=-2$ where 
the charge stiffness attains a maximum value as shown in figures 4
and 5. The existence of a characteristic peak in $D_c$
was reported in our earlier paper for $N \le 13$. Simulation results
presented here confirms our earlier results for bigger 
size systems.
What is new here is the fact that by computing the charge exponent 
$K_{\rho}$, we are now able to correlate the regime characterized by a
clear peak with the regime where the model exhibits SC fluctuations.
This also provides a plausible argument to justify the use
of an effective mean field approach (like the one
use in section III) for the Copper like
quasiparticles. Since the MF phase diagram exhibits a critical
behavior with fractal self-similar wave function, we speculate that
the phase diagram in the vicinity of the characteristic peak with strong
SC fluctuations is a dressed up version of the critical phase in
systems with strong fermion correlations.
We would like to stress that we view our MF results as providing only
qualitative predictions as the phase boundary (C-L) predicted by MF
does not agree with the one obtain by the exact diagonalization method.

Unlike the MF case where the intermediate nature of the phase
was due to the fractal nature of the single particle wave function, 
the many body phase diagram is characterized by its transport
characteristics.
Figures 4 and 5 indicate the possible existence
of a region where $D_c$ may take intermediate values:
between those of a metallic
and those corresponding to the
Anderson localized insulating phase.
Our simulations show that
the height of the peak in $D_c$ decreases rather slowly with the size
of the system. This is in contrast to  
the Anderson localized phase where
the $D_c$ decays exponentially with the size of the system.
We conjecture that in
the regime near the peak, the charge stiffness decays as a power
law. This conjecture was verified only at a 
at the special point $V=0$.
For arbitrary value of $V$, it is rather difficult to verify this
conjecture for large $N$.

\section{Conclusions and Discussion}

Two central results of this paper are: the existence of SC fluctuations
in the strongly correlated fermion model with competing length scales where the
incommensurability enhances the SC fluctuation
and a possibility of an intermediate phase (in-between metallic and
Anderson localized) which may be  related to
the critical phase of noninteracting model which exhibits
fractal characteristics. This is the first paper where the
effects of two incommensurate lengths are studied in strongly
correlated system. We hope that our results will
stimulate further studies of this type of behavior in
other models
such as $t-J$ and Hubbard models.

Our numerical calculation of the charge exponent parameter describing
SC pairing coherence provides a justification of our mean field
ideas where fermion-fermion correlations were explicitly assumed.
We want to emphasize that in aperiodic case, one cannot disregard the
mean field results in 1D as one does in the pure models. This is because,
in the pure model, the metal-insulator transition is 
the Mott transition
where a conducting phase becomes insulating due to the 
opening of a gap.
Previous studies have shown that mean field theory contradicts the
exact results as far as the existence of a gap is concerned.
In the aperiodic cases, the metal-insulator transition is 
the Anderson transition.
In the noninteracting model, this transition is characterized by the
localization of the single-particle wave function and a point spectrum.
Therefore, we would like to argue that even though mean field results are
incorrect regarding the existence or nonexistence of a gap, it may still
be of some validity in describing the Anderson transition.

Recently, there have been 
density matrix renormalization group (DMRG)
studies of spinless fermion models
that show the existence of a delocalized phase for Anderson
disordered potentials.\cite{Schmitteckert} The range of the 
parameter $V$ for which this phase seems to exist
coincides with the values of $V$ for which 
we postulated the possibility of an intermediate phase.
We think that these findings further support our
argument that an intermediate phase may exist for the
case of quasiperiodic potentials.
 
Our previous results,
hinting a new mechanism involving some sort of
competition other than the known screening of the disorder
due to SC fluctuations, are strengthened by our studies with
bigger size systems. Furthermore, the study, involving mean field 
and exact
diagonalization, provides a more convincing 
argument that this new phase may be related
to the $critical$ phase of non-interacting systems.

\acknowledgements

The research of I.I.S. is supported by a grant from National Science Foundation
DMR~097535. We also acknowledge the support of the Pittsburgh
Supercomputing Center where part of our numerical
results were obtained.
J.C.C. would like to thank the people at CIF
(International Physics Center) in Bogot\'{a}
for their constant help and encouragement during his career
and acknowledge the support obtained through
COLCIENCIAS-IDB-ICETEX from Colombia.

\begin{figure}
\caption{Self-consistent mean-field $\Delta$ vs $V$
for several values of $\lambda$. Namely,
solid line $\lambda=0.5$, dotted line $\lambda=1$,
short-dashed line $\lambda=1.5$
and long-dashed line $\lambda=2.1$.}
\label{fig1}
\end{figure}

\begin{figure}
\caption{Mean-field phase diagram in $V-\lambda$ plane showing
the E, the C and the L phases. Along the dashed
line describing the C-L transition, the model 
is conformally invariant.}
\label{fig2}
\end{figure}

\begin{figure}
\caption{The Charge exponent $K_{\rho}$ vs $V$ 
for $\lambda=0.0$ (dashed curve) and $\lambda=1.0$ (un-dashed curve).
Part (a) for $\rho = \frac{5}{13}$ and
part (b) for $\rho = \frac{13}{18}$ showing that incommensurability
enhances the SC pairing.}
\label{fig3}
\end{figure}

\begin{figure}
\caption{
Charge stiffness $D_c$
versus $V$
for $\lambda=1.0$. Part (a) shows $\rho = \frac{5}{13}$
part (b) $\rho = \frac{13}{18}$. Shaded parts indicate 
the regime where SC fluctuations exist. }
\label{fig4}
\end{figure}

\begin{figure}
\caption{
Charge stiffness
versus $V$ for $\rho = \frac{13}{21}$.
and $\lambda=1.0$. The points obtained for this
case are indicated by crosses.  The interpolated 
dashed curved is mean to be used as a guide to the eye.}
\label{fig5}
\end{figure}

\end{document}